\documentclass[10pt, conference, compsocconf]{IEEEtran}
\let\labelindent\relax

\usepackage[british]{babel}
\usepackage{xcolor,colortbl}
\usepackage{enumerate}
\usepackage{enumitem}
\usepackage{amsmath}
\usepackage{amssymb}
\usepackage{nicefrac}
\usepackage{booktabs}
\usepackage{balance}
\usepackage{wrapfig}
\usepackage{tikz}
\usepackage{pgfplots}
\pgfplotsset{compat=1.3}
\usepackage{filecontents}
\usetikzlibrary{matrix}
\usetikzlibrary{calc}
\usetikzlibrary{arrows}
\usetikzlibrary{patterns}
\usetikzlibrary{plotmarks}
\usetikzlibrary{intersections} 
\usetikzlibrary{decorations.pathmorphing}
\usetikzlibrary{decorations.pathreplacing}
\usetikzlibrary{decorations.markings,arrows} 
\usetikzlibrary{decorations.shapes}
\usetikzlibrary{fit}
\usetikzlibrary{backgrounds}
\usetikzlibrary{trees}

\let\oldhat\hat
\renewcommand{\vec}[1]{\mathbf{#1}}
\renewcommand{\hat}[1]{\oldhat{\mathbf{#1}}}

\usepackage{color}

\begin{document}
%
\title{Learning Semantic Similarity for Very Short Texts}

\author{\IEEEauthorblockN{Cedric De Boom, Steven Van Canneyt, Steven Bohez, Thomas Demeester, Bart Dhoedt}
\IEEEauthorblockA{Ghent University -- iMinds\\
Gaston Crommenlaan 8-201, 9050 Ghent, Belgium\\
\{cedric.deboom, steven.vancanneyt, steven.bohez, thomas.demeester, bart.dhoedt\}@ugent.be}
}


%


\maketitle

\begin{abstract}
Levering data on social media, such as Twitter and Facebook, requires information retrieval algorithms to become able to relate very short text fragments to each other.
Traditional text similarity methods such as tf-idf cosine-similarity, based on word overlap, mostly fail to produce good results in this case, since word overlap is little or non-existent.
Recently, distributed word representations, or word embeddings, have been shown to successfully allow words to match on the semantic level.
In order to pair short text fragments---as a concatenation of separate words---an adequate distributed sentence representation is needed, in existing literature often obtained by naively combining the individual word representations.
We therefore investigated several text representations as a combination of word embeddings in the context of semantic pair matching.
This paper investigates the effectiveness of several such naive techniques, as well as traditional tf-idf similarity, for fragments of different lengths.
Our main contribution is a first step towards a hybrid method that combines the strength of dense distributed representations---as opposed to sparse term matching---with the strength of tf-idf based methods to automatically reduce the impact of less informative terms.
Our new approach outperforms the existing techniques in a toy experimental set-up, leading to the conclusion that the combination of word embeddings and tf-idf information might lead to a better model for semantic content within very short text fragments.
\end{abstract}


%
\IEEEpeerreviewmaketitle

\section{Introduction}
On social media billions of small text messages are made public every day: own research indicates that almost every tweet is comprised of one up to approximately thirty words.
To tap into this stream of extremely short text fragments, we need appropriate information retrieval algorithms.
Tf-idf is an example of a traditional and very popular representation to compare texts, such as news articles, with each other \cite{Manning:2009uf, Achananuparp:2008uf}.
It relies on word overlap to find similarities, but in very short texts, in which word overlap is rare, tf-idf often fails.
For this reason we need sentence representations that grasp more than just word contents.

In 2013, Mikolov et al.~published three papers on the topic of distributed word embeddings to catch semantic similarities between words \cite{Mikolov:2013wl, Mikolov:2013uz, Mikolov:2013wc}, which resulted in Google's \emph{word2vec} software.
Since then scientists have extensively used such embeddings to improve state-of-the-art algorithms in natural language processing, such as part-of-speech tagging \cite{Godin:2014un}, sentence completion \cite{Hu:2014uo}, hashtag prediction \cite{Weston:2014tb}, etc.
There is however a lack of research and insight on how to effectively combine embeddings into a single sentence representation that contains most of its semantic information.
Many authors choose to average or maximize across the embeddings in a text \cite{Weston:2014tb, dosSantos:2014tr, Collobert:2011tk} or combine them through a multi-layer perceptron \cite{Godin:2014un, Kang:2014un}, by clustering \cite{Zhang:2015wt}, or by trimming the text to a fixed length \cite{Kang:2014un}.

The Paragraph Vector algorithm by Le and Mikolov---also termed \emph{paragraph2vec}---is a powerful method to find suitable vector representations for sentences, paragraphs and documents of variable length \cite{Le:2014vd}. The algorithm tries to find embeddings for separate words and paragraphs at the same time through a procedure similar to word2vec. The collection of paragraphs, however, is known beforehand. This implies that finding a vector representation for a new and probably unseen paragraph---the theoretical number of different paragraphs is after all many times higher than the number of different words---requires additional training. Paragraph2vec is therefore not a fit candidate to be used in, e.g., a stream of messages as is the case with social media.

Further research is thus needed to derive optimal sentence representations based on word embeddings. By investigating and comparing the performance of several word combination approaches in a short-text matching task, we arrive at a novel technique in which we aggregate both tf-idf and word embedding signals.
In this paper, we show how word embeddings can be combined into a new vector representation for the entire considered fragment, in which the impact of frequent words---i.e.~with a low idf-component, and therefore mostly non-informative---is reduced with respect to more informative words.
This leads to a significant increase in the effectiveness of detecting semantically similar short-text fragments, compared to traditional tf-idf techniques or simple heuristic methods to combine word embeddings.
Our approach is a first step towards a hybrid method that unites word embedding and tf-idf information of a short text fragment into a distributed representation that catches most of that fragment's semantic information.

Very recently, Kusner et al.~devised a simple method to measure the similarity between documents based on the minimal distance word embeddings have to travel from one document to another \cite{Kusner:2015vc}. In this, the authors only consider non stop words, and evaluate their distance measure using $k$NN classification. We, however, will learn vector representations for documents, and we will evaluate our technique through a newly crafted dataset of related text fragments. Also recently, Zheng and Callan created a simple algorithm based on linear regression to find relevant terms in a query \cite{Zheng:2015kz}. The authors learned weights for each dimension in a word embedding using a supervised relevance signal. Our work is different in that we will learn weights for entire word vectors instead of separate dimensions. For this we will use tf-idf information, instead of only word embedding features. Furthermore, we will arrive at a globally applicable weighting scheme instead of a query-dependent one.

In the next section we will discuss our experimental setup and method of data collection, and explain how well a number of traditional techniques perform on our dataset.
We will then use the gained insights to create more effective distributed representations by integration of the tf-idf information.

\section{Experimental Set-up and Analysis}
To evaluate techniques that measure semantic similarity between short text fragments, we need a reference set with couples of fragments that are semantically related, and couples which are not related. 
We denote the former as a \emph{pair}, and the latter as a \emph{non-pair}.
Every couple consists of two texts, which are built up as a sequence of words.
For an arbitrary couple we introduce the notation $c$, and the two texts in $c$ are denoted as the sequences $(c^{1})$ and $(c^{2})$.
Element $j$ of sequence $(c^{1})$ is the vector of word $j$ in the first text of $c$, denoted as $\vec{w}^{1}_j$:
\begin{align*}
\left(c^{1}\right) \triangleq \left(\vec{w}^{1}_1, \vec{w}^{1}_2, \vec{w}^{1}_3, \dots\right).
\end{align*}
A vector representation for $(c^{1})$, combining the word representations contained in $(c^{1})$, is written as $\vec{o}^{1}$, and as $\vec{o}^{2}$ for $(c^{2})$ respectively.

In this paper we strive to give the initial impetus to learning sentence representations of very short text fragments mainly found on social media, but for now we will perform our experiments in a toy environment using English Wikipedia articles.
These are of course very different textual media---which has some disadvantages, as we will discuss later---but Wikipedia articles have the benefit of being well-structured, which allows us to extract related texts more easily.
In our experiments we use the Wikipedia dump of March 4th, 2015, after cleaning the articles by removing markup and punctuation.
We convert all texts to lowercase and replace the numbers by a single character `0'.
In our toy setting we require that the texts of all couples are composed of the same number of words, i.e.~the length of the sequences $(c^{1})$ and $(c^{2})$ is equal to $n_c$.
To extract a pair of texts, each containing $n_c$ words, we take the first $n_c$ words of a paragraph, skip the next two words, and then take the following $n_c$ words.
To extract a non-pair, we take $n_c$ words out of two random paragraphs of different articles.
This approach is closely related to the one used by Hu et al.~to extract pairs and non-pairs from the Reuters corpus \cite{Hu:2014uo}.
In total we extract five million pairs and five million non-pairs, for texts of ten, twenty, and thirty words long\footnote{The dataset can be obtained upon request.}.

To represent words as a vector, we train word embeddings on the entire Wikipedia corpus.
We do this through Google's \emph{word2vec} software, using skip-gram with negative sampling, a context window of five words, and 400 dimensions.
We also calculate document frequencies for every word using the same Wikipedia corpus.

We regard two text fragments to be semantically similar if their corresponding vector representations lie close to each other according to some distance measure, and dissimilar if the vectors lie farther apart.
Semantic similarity between text fragments is therefore related to semantic similarity between skip-gram word embeddings, in which the cosine distance between related words is smaller compared to unrelated words.
This is also the reason why we do not use paraphrase datasets, such as the Microsoft Research Paraphrase corpus or the SemEval2015 Twitter Paraphrase dataset, to perform our experiments. After all, the notion of semantic relatedness in these datasets is often too narrow: if one sentence is about Star Wars and another about Anakin Skywalker, they are semantically related although they might not be paraphrases of each other.

To verify whether our own dataset of 10 million Wikipedia couples is a valid candidate to perform similarity experiments on, we will test different techniques that try to enhance the discriminative power between pairs and non-pairs as much as possible---i.e.~leading to small distances between pairs and larger distances between non-pairs.
We start with techniques ranging from plain tf-idf to naive combinations of word embeddings, after which we investigate elementary mixtures of tf-idf and word embedding signals.

\begin{figure}[t!]
\centering
\begin{tikzpicture}[scale=.84,inner sep=0pt]
	\node [draw=none, anchor=south west] () at (0,0) {\includegraphics[trim = 2.5cm 1.5cm 1.5cm 1.5cm, clip, width=.79\linewidth]{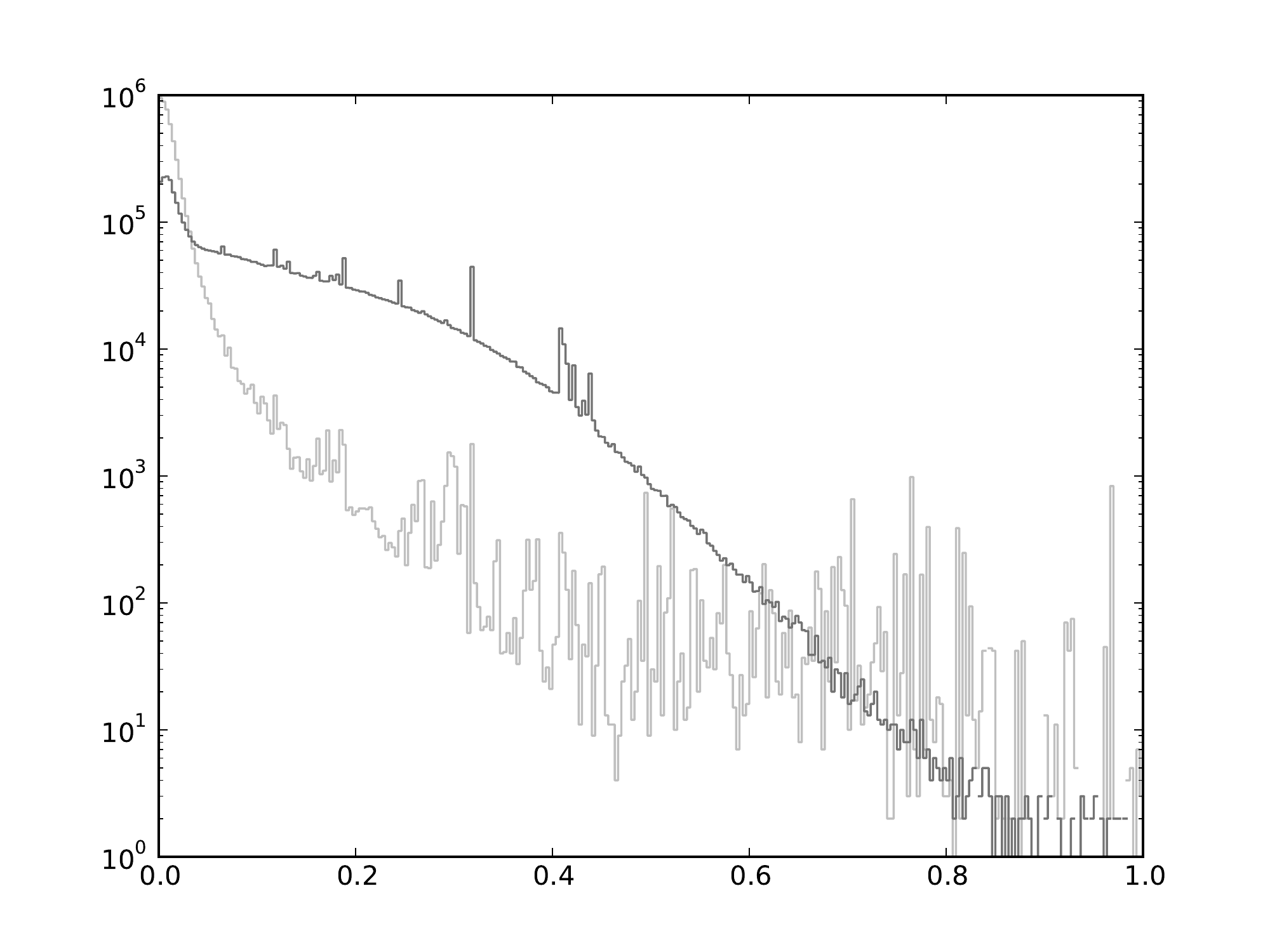}};
	
	\coordinate (TL) at (0,6.0);
	\coordinate (TR) at (7.8,6.0);
	\coordinate (BL) at (0,0);
	\coordinate (BR) at (7.8,0);
	
	\node[below=2pt] (bO) at (BL) {$0.0$};
	\node[below=2pt] (b1) at (1.56,0.0) {$0.2$};
	\node[below=2pt] (b2) at (3.12,0.0) {$0.4$};
	\node[below=2pt] (b3) at (4.68,0.0) {$0.6$};
	\node[below=2pt] (b4) at (6.24,0.0) {$0.8$};
	\node[below=2pt] (b5) at (7.8,0.0) {$1.0$};
	
	\draw[draw opacity=0] (BL) -- node[below=12pt] {Cosine similarity} (BR);
	\draw[draw opacity=0] (BL) -- node[left=18pt] {\rotatebox{90}{Number of couples}} (TL);
	\node[left=2pt] (lO) at (BL) {$10^0$};
	\node[left=2pt] (l1) at (0.0,1.0) {$10^1$};
	\node[left=2pt] (l2) at (0.0,2.0) {$10^2$};
	\node[left=2pt] (l3) at (0.0,3.0) {$10^3$};
	\node[left=2pt] (l4) at (0.0,4.0) {$10^4$};
	\node[left=2pt] (l5) at (0.0,5.0) {$10^5$};
	\node[left=2pt] (l6) at (0.0,6.0) {$10^6$};
\end{tikzpicture}
\caption{Histogram plot of the number of couples as a function of their cosine similarity using tf-idf, for both pairs (dark grey) and non-pairs (light grey).
}
\label{fig:tfidf_plot}
\end{figure}
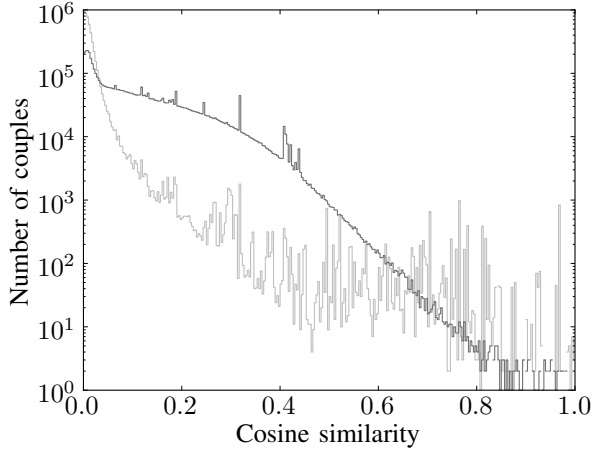

For every couple $c$ we create a tf-idf representation for both $(c^{1})$ and $(c^{2})$, and calculate the cosine similarity between $(c^{1})$ and $(c^{2})$.
Figure \ref{fig:tfidf_plot} shows a histogram plot of the number of couples as a function of their cosine similarity, for both pairs and non-pairs separately, and for texts of 20 words long.
We see that there are many couples having a very low cosine similarity, which is due to the very short length of the text fragments.
There are many more pairs having a larger cosine similarity than there are non-pairs, but non-related texts can also exhibit relatively large similarity values, which is due to coincidental overlap of mostly non-informative words.

As for word embeddings, we create two traditional sentence representations as a baseline.
In a first representation we take the mean of all word embeddings in the text:
\begin{align}
\forall \ell \in \{1,2\}\colon \vec{o}^{\ell} = \frac{1}{n_c} \sum_{j=1}^{n_c} \vec{w}^{\ell}_j,
\end{align}
in which $\vec{w}^{\ell}_j$ represents the \textit{word2vec} word embedding vector of the $j$'th word in text sequence $\ell$.
In a second representation, we take for each dimension the maximum across all embeddings:
\begin{align}
\forall \ell \in \{1,2\}, k\in \{1,\dots, 400\}\colon \vec{o}^{\ell}_k = \max_{j} \vec{w}^{\ell}_{j,k}.
\end{align}
Between two such representations we then calculate the cosine similarity as before.

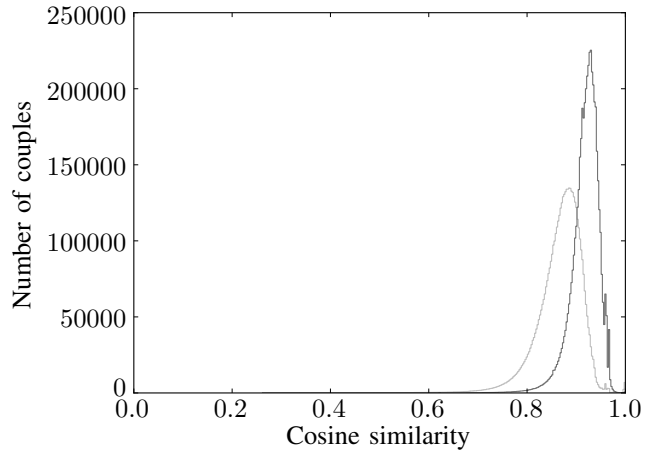
\begin{figure}[t!]
\centering
\begin{tikzpicture}[scale=.84,inner sep=0pt]
	\node [draw=none, anchor=south west] () at (0,0) {\includegraphics[trim = 2.5cm 1.5cm 1.5cm 1.5cm, clip, width=.79\linewidth]{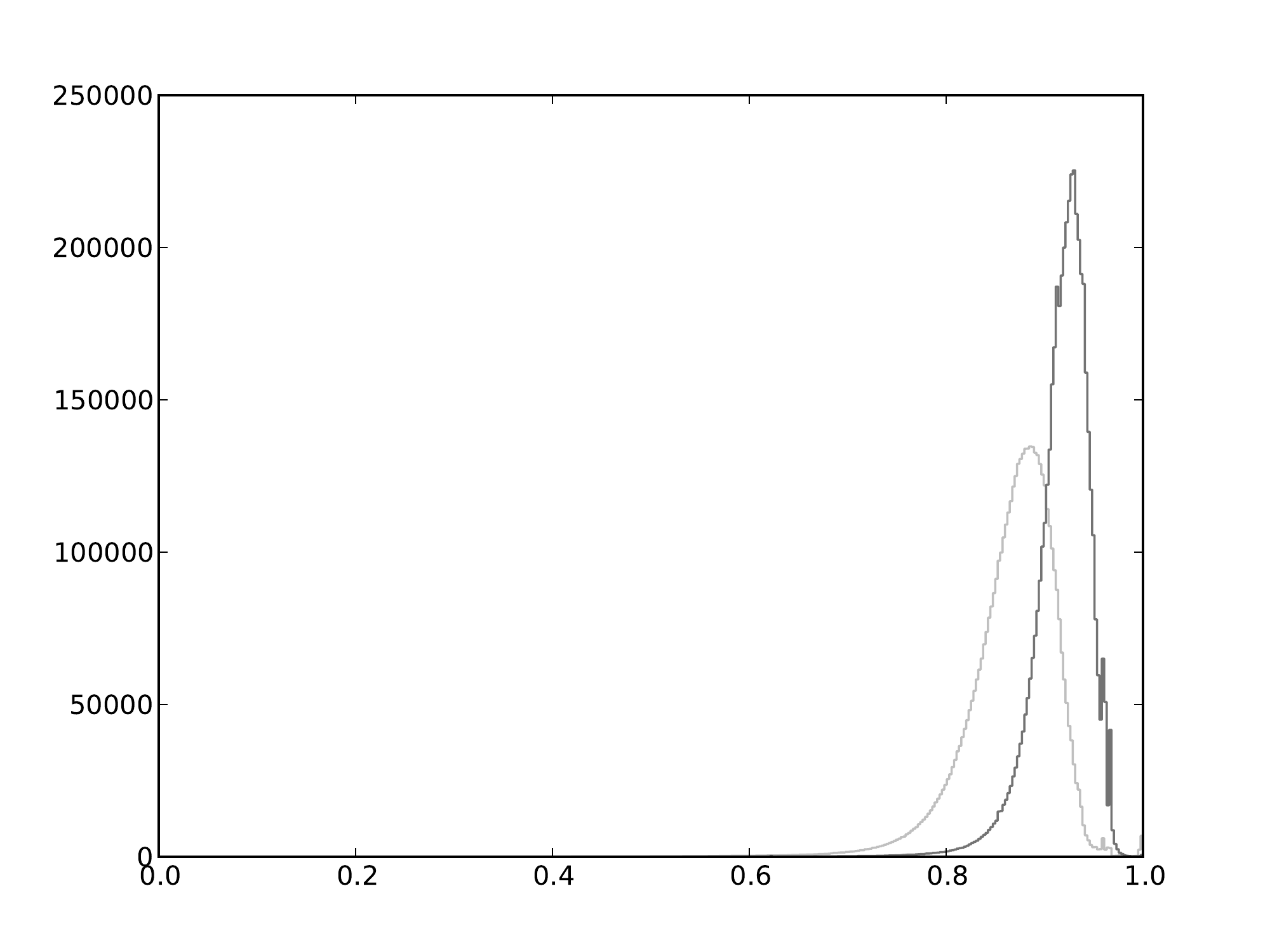}};
	
	\coordinate (TL) at (0,6.0);
	\coordinate (TR) at (7.8,6.0);
	\coordinate (BL) at (0,0);
	\coordinate (BR) at (7.8,0);
	
	\node[below=2pt] (bO) at (BL) {$0.0$};
	\node[below=2pt] (b1) at (1.56,0.0) {$0.2$};
	\node[below=2pt] (b2) at (3.12,0.0) {$0.4$};
	\node[below=2pt] (b3) at (4.68,0.0) {$0.6$};
	\node[below=2pt] (b4) at (6.24,0.0) {$0.8$};
	\node[below=2pt] (b5) at (7.8,0.0) {$1.0$};
	
	\draw[draw opacity=0] (BL) -- node[below=12pt] {Cosine similarity} (BR);
	\draw[draw opacity=0] (BL) -- node[left=36pt] {\rotatebox{90}{Number of couples}} (TL);
	\node[left=2pt] (lO) at (0.0,0.1) {$0$};
	\node[left=2pt] (l1) at (0.0,1.2) {$50000$};
	\node[left=2pt] (l2) at (0.0,2.4) {$100000$};
	\node[left=2pt] (l3) at (0.0,3.6) {$150000$};
	\node[left=2pt] (l4) at (0.0,4.8) {$200000$};
	\node[left=2pt] (l5) at (0.0,6.0) {$250000$};
\end{tikzpicture}
\caption{Histogram plot of the number of couples as a function of their cosine similarity using the mean of the word embeddings, for both pairs (dark grey) and non-pairs (light grey).
}
\label{fig:mean_plot_1}
\end{figure}

Figure \ref{fig:mean_plot_1} shows a histogram plot for the mean of the embeddings, and for texts of 20 words long.
The graph shows two curves with a stretched tail towards lower cosine similarities.
We see that the mode of the non-pairs curve lies more to the left than the mode of the pairs curve, but still close to each other.
Our hypothesis is that this is due to overlap in non-informative but frequently occurring words, such as articles and prepositions.
Such words, however, contribute little to the semantic meaning of a text, and by reducing the influence of such words, we want to accentuate the true semantics of a text fragment.
By reducing this coincidental similarity, we intend to shift the non-pairs stronger toward lower similarities than the pairs, hence increasing the resolution between both.

Since less informative terms are common to many sentences, they mostly have a high document frequency as well.
We therefore implement the mean and max techniques again, but this time we only use the top 30\% of the words with the highest idf component.
In a final technique we use all words, but we weigh each word vector with its idf value, after which we take the mean.

\begin{table*}
\small
\centering
\caption{Comparison of different word vector aggregation techniques with cosine similarity.}
\label{table:analysis}
\begin{tabular}{l | c c | c c | c c}
\toprule
					& \multicolumn{2}{c|}{10 words} & \multicolumn{2}{c|}{20 words} & \multicolumn{2}{c}{30 words}\\
					& Split error & JS divergence & Split error & JS divergence & Split error & JS divergence\\
\hline
Tf-idf				& 36.15\% & 0.11946 & 20.09\% & 0.35991 & 12.55\% & 0.54468\\
Mean					& 30.67\% & 0.16186 & 21.05\% & 0.34109 & 16.33\% & 0.45534\\
Max					& 28.27\% & 0.20831 & 19.06\% & 0.40030 & 15.18\% & 0.49882\\
\hline
Min					& 28.89\% & 0.19694 & 19.54\% & 0.38696 & 15.69\% & 0.48592\\
Min/max 				& 26.89\% & 0.22946 & 16.86\% & 0.45004 & 12.76\% & 0.56253\\
\hline
Mean, top 30\% idf		& 23.69\% & 0.30044 & 15.86\% & \textbf{0.48260} & \textbf{12.42\%} & \textbf{0.56456}\\
Max, top 30\% idf		& 26.56\% & 0.24028 & 20.63\% & 0.36809 & 16.66\% & 0.45522\\
Min/max, top 30\% idf	& 25.43\% & 0.25761 & 19.13\% & 0.39775 & 14.91\% & 0.49927\\
Mean, idf weighed 		& \textbf{23.35\%} & \textbf{0.30400} & \textbf{15.77\%} & 0.48028 & 12.43\% & 0.56028\\
\bottomrule
\end{tabular}
\end{table*}

As word embeddings contain both positive and negative numbers, we also test the influence of the sign in these embeddings.
In a first experiment we take, instead of the maximum, the minimum across all word embeddings in a text.
In a second experiment we test whether extremes, either positive or negative, are important indicators for semantic similarity.
We therefore simply concatenate the maximum vector and the minimum vector to form a new vector representation.

To evaluate the power of the previously described techniques to discriminate between pairs and non-pairs, we calculate two performance metrics: optimal split error and Jensen-Shannon (JS) divergence.
We obtain the former using the optimal threshold for which the number of misclassified couples is minimal.
The latter is a symmetric measure expressing the similarity between two probability distributions, based on the well-known---but asymmetric---KL divergence.
The lower the optimal split error or the higher the JS divergence, the better a technique can distinguish pairs from non-pairs.
Table \ref{table:analysis} shows the results, for texts of 10, 20 and 30 words.


As for the traditional techniques, the max approach works best for 10 and 20 words, but as the number of words increases to 30, tf-idf performs best, which is logical since word overlap rises with a growing number of words, as can be seen in Figure \ref{fig:tfidf_plot} as well.
The min approach performs almost as well as the max approach.
By concatenating the minimum and maximum vectors, we see that split error and JS divergence are improved by a large margin.
We can thus conclude that the sign in word embeddings holds complementary semantic information.
By incorporating document frequency information, we do better than all previous techniques for texts of 10 words long.
But for longer texts the mean approach performs best, while the min and max combinations achieve worse results than when using the complete text.
The best performing techniques are the idf-weighed mean approach and the approach taking the mean of 30\% of the word vectors with the highest idf-components, as they perform similarly across the different word lengths.

We also investigate the influence of the used distance metric.
We test cosine distance, Euclidean distance, $L_3$-norm, $L_4$-norm and Bray-Curtis distance, which are normalised between 0 and 1.
We use texts of 20 words long and the mean of the embeddings. Table \ref{table:distances} shows the results, again expressed in terms of split error and JS divergence.
Euclidean distances perform best in our tests, so we continue to use Euclidean distances hereafter.

\begin{table}
\small
\centering
\caption{Comparison of different distance metrics for texts of 20 words long.}
\label{table:distances}
\begin{tabular}{l | c c c}
\toprule
 					&\hspace{0.4cm} & Split error & JS divergence\\
\hline
Mean, cosine	        &   & 21.05\% 	& 0.34109\\
Mean, Euclidean		&	& \textbf{19.55\%} & \textbf{0.37788}\\
Mean, $L_3$			&	& 19.62\% & 0.37511\\
Mean, $L_4$			&	& 19.77\% & 0.37061\\
Mean, Bray-Curtis	&	& 21.22\% & 0.33775\\
\bottomrule
\end{tabular}
\end{table}

\section{Learning Semantic Similarity}
As became clear from the data analysis, combining knowledge from both tf-idf and word embeddings can be beneficial.
Using only the portion with the highest idf component of all words clearly reduces split error and improves JS divergence.
After all, low-idf words have no clear-cut semantic meaning, and since these words are present in many sentences, there is more coincidental overlap between non-related sentences. Removing these words---or lowering their influence---from a text representation thus succeeds in pulling apart the average similarity between pairs and between non-pairs.

In this section we investigate how we can learn to optimally weigh words in a short text.
This way we intend to do better than just taking the top-idf words or weighing these words with their idf component, in order to maximize the average distance between pairs and non-pairs.
As before, we perform the experiments in a toy setting on couples of short Wikipedia texts, with as many pairs as non-pairs.
We divide the total dataset into a training set $\mathcal{D}$ of 1.5 million couples, a test set $\mathcal{T}$ of 1.5 million couples and a validation set $\mathcal{V}$ of 2.0 million couples.
Since we describe ongoing research and present the first steps towards a flexible hybrid technique, we only consider texts of 20 words long in this section, and varying the fragment length will be suggested as future work.

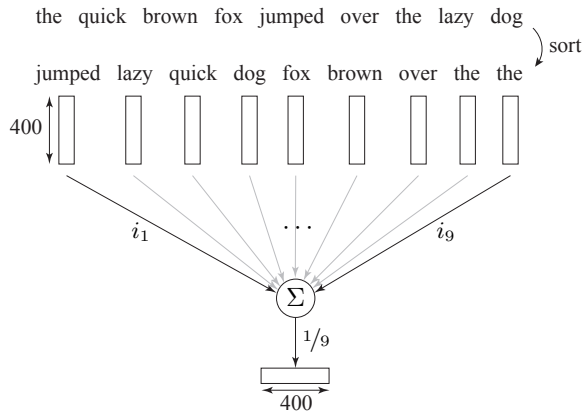
\begin{figure}[t!]
\centering
\begin{tikzpicture}[scale=0.97,inner sep=0pt]
	\node [draw=none, anchor=south west] () at (0,0) {\includegraphics[width=.9080\linewidth]{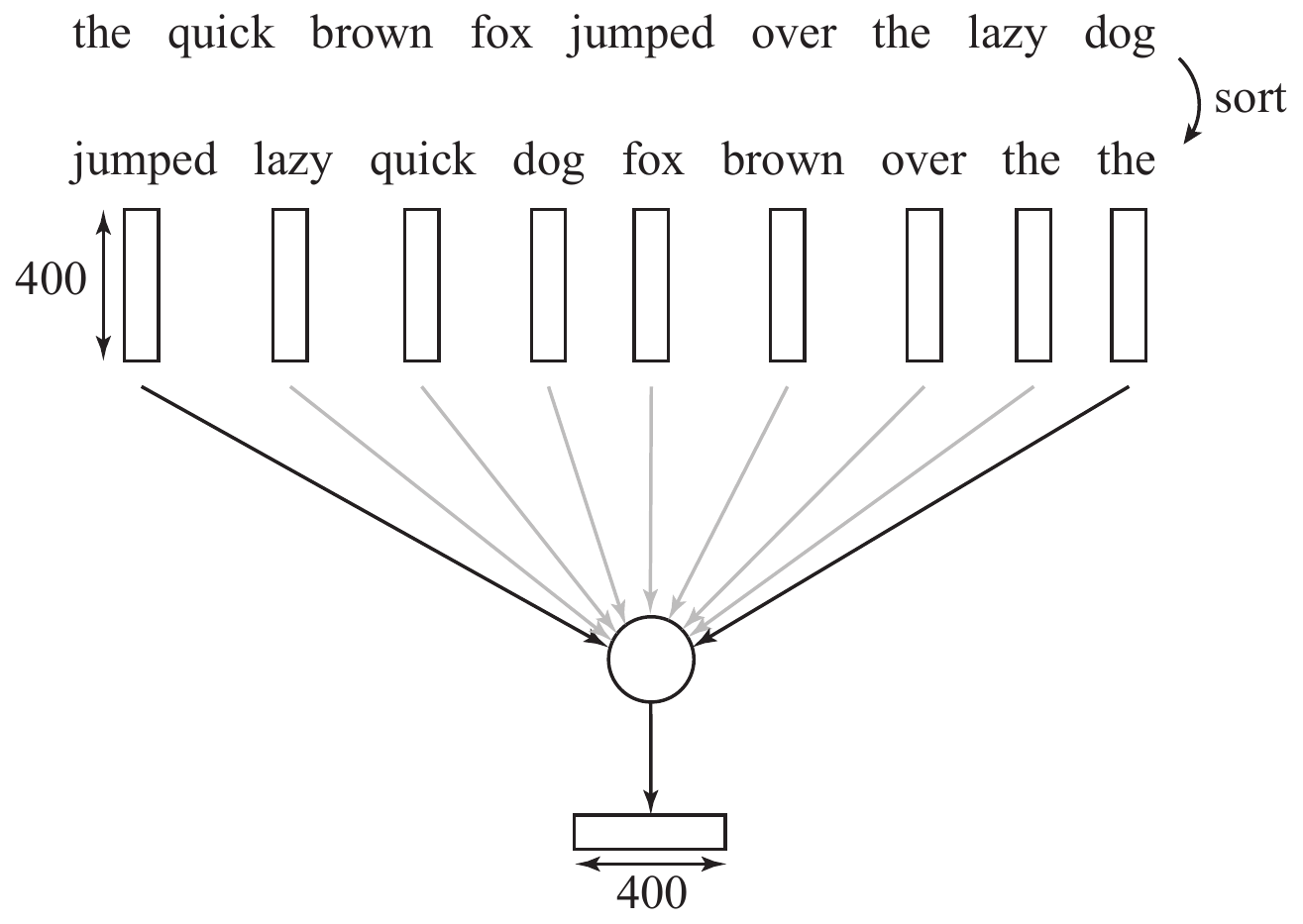}};
	
	\node[below=2pt] (bO) at (1.9,2.75) {\small $i_1$};
	\node[below=2pt] (b8) at (6.08,2.75) {\small $i_9$};
	
	\node[below=2pt] (d) at (4.08,2.75) {$\cdots$};
	
	\node[below=2pt] (s) at (4.02,1.85) {$\Sigma$};
	
	\node[below=2pt] (n) at (4.3,1.30) {\small $\nicefrac{1}{9}$};
	
\end{tikzpicture}
\caption{Illustration of the importance factor approach for a toy sentence of nine words long.}
\label{fig:example_1}
\end{figure}

We implement the following learning procedure.
For every couple $c$ in the training set we sort the words in both texts $(c^{1})$ and $(c^{2})$ according to their document frequency---i.e.~the word with the lowest document frequency comes first---arriving at $(c^{1'})$ and $(c^{2'})$.
Next we multiply the word embedding vector of each word $\vec{w}^{1'}_j$ and $\vec{w}^{2'}_j$ with an importance factor $i_j$; these importance factors are global weights that will be learned.
Finally, we take the mean of these weighed embeddings to obtain a fixed-length vector $\vec{o}^{1}$ for $(c^{1})$ and $\vec{o}^{2}$ for $(c^{2})$:
\begin{equation}
\forall \ell \in \{1,2\}\colon \vec{o}^{\ell} = \frac{1}{n_c} \sum_{j = 1}^{n_c} i_j \cdot \vec{w}^{\ell'}_j.
\end{equation}
We take the mean since it is the best performing technique in the third part of Table \ref{table:analysis}. Figure \ref{fig:example_1} illustrates the entire procedure of calculating a vector representation for a sentence using the importance factor method. We see that first the words in the sentence are sorted according to their idf-component; next, their 400-dimensional word embedding vectors are multiplied by importance factors, and finally the mean is taken.

\begin{figure}[t!]
\centering
\begin{tikzpicture}[scale=.84,inner sep=0pt]
	\node [draw=none, anchor=south west] () at (0,0) {\includegraphics[trim = 2.5cm 1.5cm 1.5cm 1.5cm, clip, width=.79\linewidth]{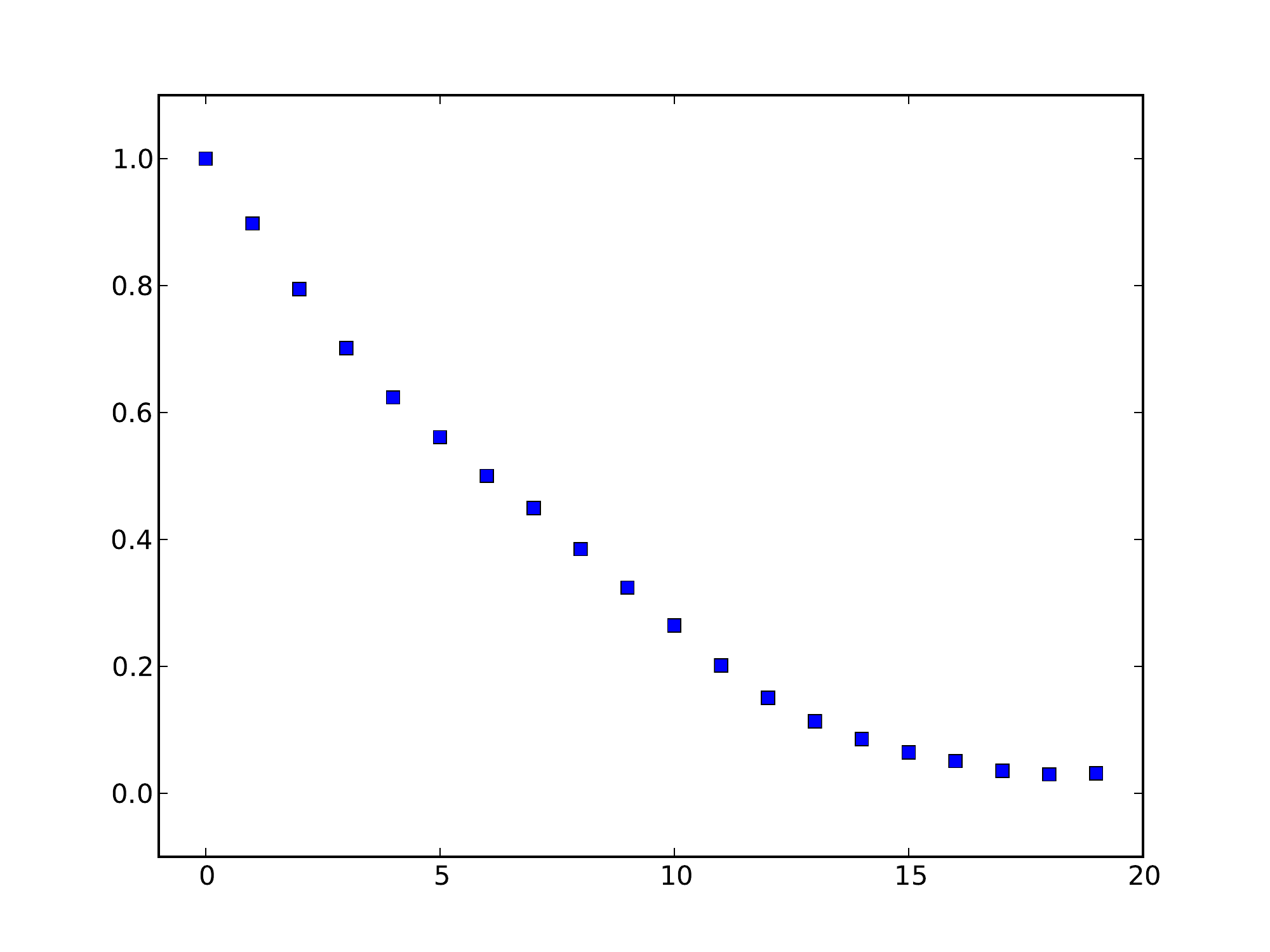}};
	
	\coordinate (TL) at (0,6.0);
	\coordinate (TR) at (7.8,6.0);
	\coordinate (BL) at (0.4,0);
	\coordinate (BR) at (7.4,0);
	
	\node[below=2pt] (bO) at (0.4,0.0) {$1$};
	\node[below=2pt] (b1) at (2.25,0.0) {$6$};
	\node[below=2pt] (b2) at (4.1,0.0) {${11}$};
	\node[below=2pt] (b3) at (5.95,0.0) {${16}$};
	\draw[draw opacity=0] (BL) -- node[below=12pt] {Importance factor index $j$} (BR);
	\draw[draw opacity=0] (0.0, 0.5) -- node[left=3pt] {\rotatebox{90}{Magnitude $i_j$}} (0.0, 5.5);
	\node[left=2pt] (lO) at (0.0,0.5) {$0$};
	\node[left=2pt] (l1) at (0.0,5.5) {$1$};
\end{tikzpicture}
\caption{Plot of the importance factor magnitudes.}
\label{fig:importance_factors}
\end{figure}
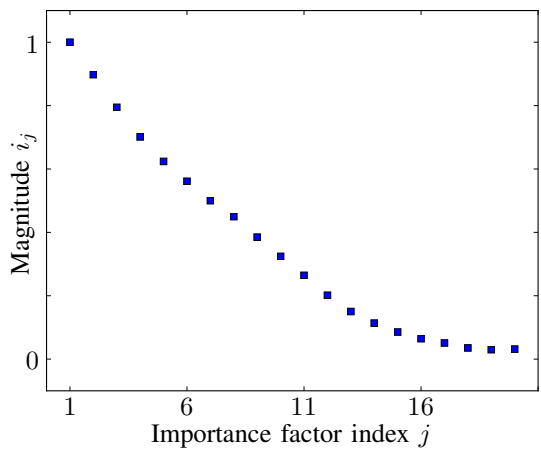

To learn the importance factors, we define a loss function as a function of any couple $c$ that minimizes the distance between the vectors of a pair, and maximizes the distance between the vectors of a non-pair:
\begin{align}
f(c) \triangleq \begin{cases} d(\vec{o}^{1}, \vec{o}^{2}) & \text{if } c\text{ is a pair} \\ - d(\vec{o}^{1}, \vec{o}^{2}) & \text{if } c\text{ is a non-pair} \end{cases}
\end{align}
with $d(\cdot)$ a distance function of choice.
We use a squared Euclidean distance as distance function:
\begin{align}
d(\vec{o}^{1}, \vec{o}^{2}) = \sum_{j=1}^{n_c} (\vec{o}^{1}_j - \vec{o}^{2}_j)^2.
\end{align}
We then optimize the following objective as a function of the importance factors:
\begin{align}
J(i_1,\dots,i_{n_c}) = \frac{1}{|\mathcal{D}|} \sum_{c\in\mathcal{D}} f(c) + \lambda\sum_{j=1}^{n_c} i_j^2.
\end{align}
To minimize this objective function we use stochastic gradient descent with batches of 100 couples, a learning rate of 0.1, a momentum of 0.9, and a regularization constant $\lambda$ of 0.0015. We start the optimization with all importance factors equal to 0.5.
Thanks to the large amount of couples in the training set, we can stop the optimization after one epoch of training.
By then, the factors have settled to an optimum and the procedure has seen all training couples exactly once, thereby reducing chances of overfitting.

Figure \ref{fig:importance_factors} shows a plot of the importance factors that were learned through the earlier-described optimization procedure.
We clearly notice that the importance factors steadily decrease in magnitude; words with a low document frequency therefore weigh much more than words with a high document frequency, which confirms our hypothesis.
The factors at the end are very close to zero.

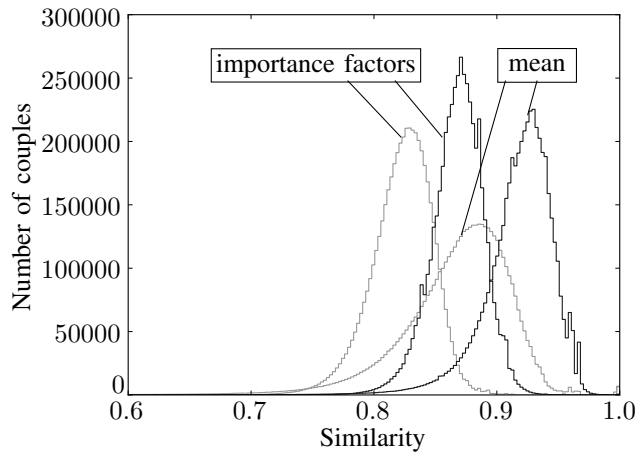
\begin{figure}[t!]
\centering
\begin{tikzpicture}[scale=.84,inner sep=0pt]
	\node [draw=none, anchor=south west] () at (0,0) {\includegraphics[trim = 2.5cm 1.5cm 1.5cm 1.5cm, clip, width=.79\linewidth]{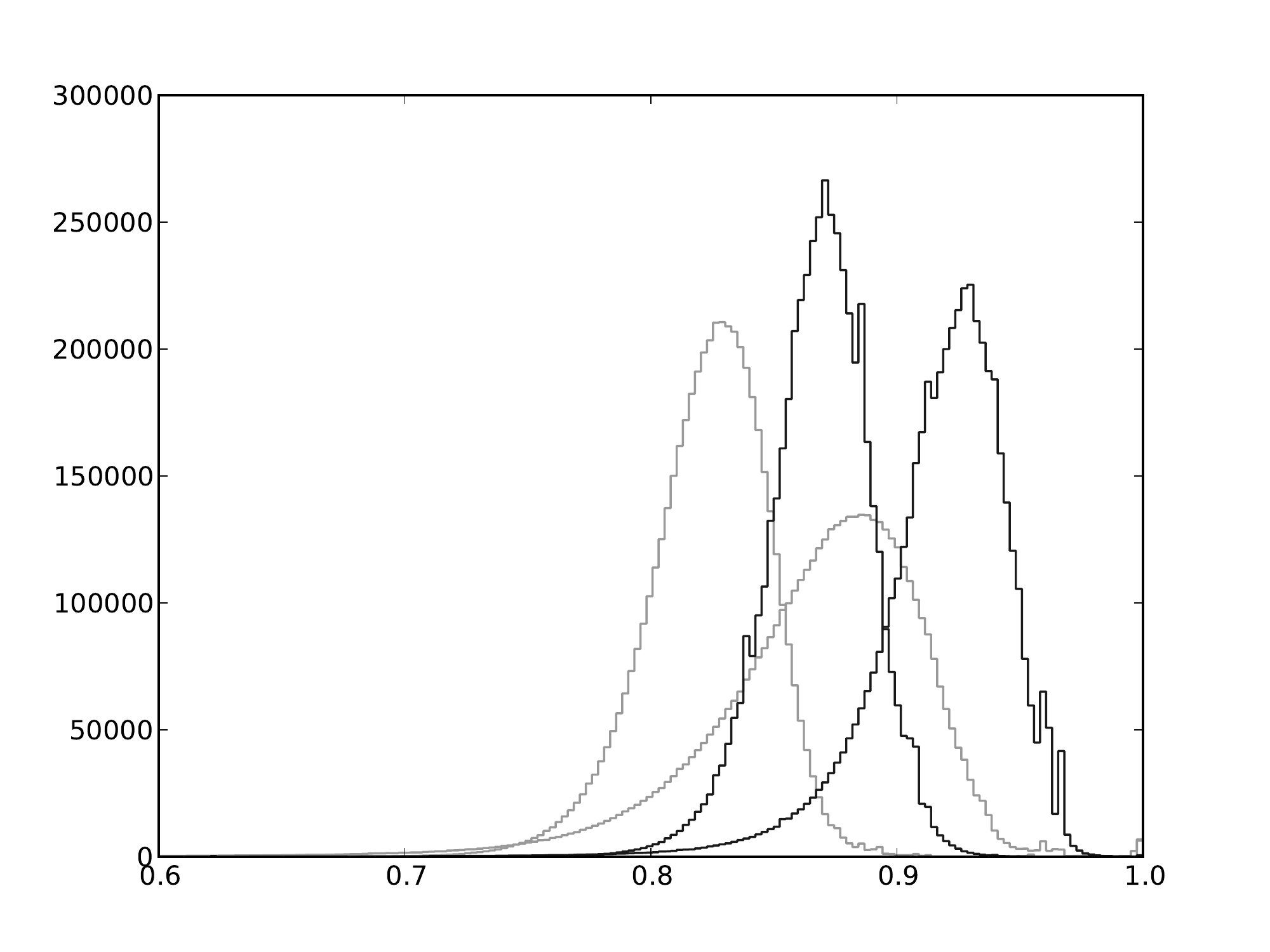}};
	
	\coordinate (TL) at (0,6.0);
	\coordinate (TR) at (7.8,6.0);
	\coordinate (BL) at (0.0,0);
	\coordinate (BR) at (7.8,0);
	
	\node[below=2pt] (b0) at (0.0,0.0) {$0.6$};
	\node[below=2pt] (b1) at (1.95,0.0) {$0.7$};
	\node[below=2pt] (b2) at (3.9,0.0) {$0.8$};
	\node[below=2pt] (b3) at (5.85,0.0) {$0.9$};
	\node[below=2pt] (b4) at (7.8,0.0) {$1.0$};
	\draw[draw opacity=0] (BL) -- node[below=12pt] {Similarity} (BR);
	\draw[draw opacity=0] (BL) -- node[left=34pt] {\rotatebox{90}{Number of couples}} (TL);
	\node[left=3pt, above=1pt] (lO) at (BL) {$0$};
	\node[left=2pt] (l1) at (0.0,1.0) {$50000$};
	\node[left=2pt] (l2) at (0.0,2.0) {$100000$};
	\node[left=2pt] (l3) at (0.0,3.0) {$150000$};
	\node[left=2pt] (l4) at (0.0,4.0) {$200000$};
	\node[left=2pt] (l5) at (0.0,5.0) {$250000$};
	\node[left=2pt] (l6) at (0.0,6.0) {$300000$};
	
	\draw[color=black] (4.35, 4.1) -- (3.5,5.0) {};
	\draw[color=black] (5.0, 4.1) -- (4.25,5.0) {};
	\draw (3.0,5.24) node 
        [rectangle, draw, align=center, inner sep=0.4ex] {importance factors};
        
     \draw[color=black] (5.3, 2.55) -- (6.0,5.0) {};
     \draw[color=black] (6.35, 4.45) -- (6.5,5.0) {};
     \draw (6.5,5.24) node 
        [rectangle, draw, align=center, inner sep=0.9ex] {mean};
\end{tikzpicture}
\caption{Comparison between mean embeddings and the importance factor approach, for both pairs (dark grey) and non-pairs (light grey).}
\label{fig:mean_plot_2}
\end{figure}

To compare the performance of our importance factor approach to the earlier-described combination techniques, we calculate the optimal split point between pairs and non-pairs on our validation set for each of the techniques.
Using these optimal split points, we then calculate the final split error rate on our test set.
As a distance metric we use a normalized Euclidean distance, except for tf-idf for which we used the standard cosine distance.
Table \ref{table:testset_eucl_errorrate} shows the error rates for the different techniques we tested.
We notice that the importance factor technique outperforms the other approaches by approx 2.0\% in error rate.
This is a significant decrease, as $p < 0.001$ in a two-tailed binomial test.
We compare our importance factor approach with the plain mean technique on the complete dataset in Figure \ref{fig:mean_plot_2}, in which we see that in our approach the variance of the curves is smaller and that there is less overlap between the pairs and non-pairs.

\section{Discussion and Conclusion}
We gained insight in the power of tf-idf and several word embedding aggregation techniques to relate pairs of very short texts to each other.
We also learned how to optimally combine knowledge from both tf-idf and word embeddings to maximize the separation between pairs and non-pairs.
The best performing traditional technique is a concatenation of maximum and minimum vectors, and for a large number of words tf-idf produces comparable results.
Our importance factor approach, however, significantly outperforms all other techniques by a large margin.

\begin{table}
\small
\centering
\caption{Comparison of error rates on the test set, indicating a significant reduction for the importance factor approach.}
\label{table:testset_eucl_errorrate}
\begin{tabular}{l | c c c }
\toprule
						& \hspace{0.4cm} & Error rate & \hspace{0.4cm}\\
\hline
Tf-idf					& & 19.60\% & \\
Mean						& & 19.43\% & \\
Max						& & 19.05\% & \\
Min/max					& & 16.78\% & \\
Mean, top 30\% idf		& & 17.02\% & \\
Max, top 30\% idf		& & 18.05\% & \\
Min/max, top 30\% idf	& & 16.40\% & \\
Mean, idf weighed 		& & 24.00\% & \\
Mean, importance factors & & \textbf{14.44\%}\\
\bottomrule
\end{tabular}
\end{table}

In this paper we have laid out the first steps towards a flexible and hybrid technique to combine word embeddings with tf-idf information. Since this is still ongoing research, a few remarks need, however, to be made. First, our experimental set-up is close to a toy setting. Wikipedia is a completely different textual medium than a social platform such as Twitter, in which the used language is full of slang, hashtags and spelling errors. In future work we will therefore adapt our novel technique to texts found in social media posts.

Secondly, our current approach is limited to texts of a fixed length, while all other techniques in Table \ref{table:testset_eucl_errorrate} do not suffer from this restriction. This forms a strong constraint on the applicability of the technique. However, current research shows promise that the approach can be extended to texts of arbitrary length as well, but this still needs to be investigated further in future work.

In other future work we will discuss yet other combination schemes of word embeddings and tf-idf, and experiment with the number of dimensions in the embeddings. We will also investigate and compare the performance of LSI, topic models such as LDA, and other state-of-the-art document distance measures based on word embeddings.

\section{Acknowledgments}
Cedric De Boom is funded by a Ph.D.~grant of Ghent University, Special Research Fund (BOF) and of the Flanders Research Foundation (FWO).\\
Steven Van Canneyt and Steven Bohez are funded by a Ph.D.~grant of the Agency for Innovation by Science and Technology in Flanders (IWT).

\balance



\bibliographystyle{IEEEtran}
\bibliography{lib}

\end{document}